\documentclass[proof]{WileyASNA-v1}

\usepackage{amsmath}
\usepackage{graphicx}
\usepackage{siunitx} 
\usepackage{nicefrac} 
\usepackage{commath} 

\usepackage{mathtools} 

\DeclarePairedDelimiter\floor{\lfloor}{\rfloor}

\makeatletter
\g@addto@macro\@floatboxreset\centering
\makeatother

\articletype{Original Article}%

\received{\today}
\revised{tbc}
\accepted{tbc}

\begin{document}

\title{Transfer Learning in Spatial-Temporal Forecasting of the Solar Magnetic Field}

\author{Eurico Covas}

\authormark{Covas}

\address{\orgdiv{CITEUC — Center for Earth and Space Science Research of the University of Coimbra}, \orgname{Geophysical and Astronomical Observatory of the University of Coimbra}, \orgaddress{\country{3040-004 Coimbra, Portugal}}}


\corres{Eurico Covas: \email{eurico.covas@mail.com}}

\fundingInfo{CITEUC is funded by National Funds through FCT - Foundation for Science
and Technology (project: UID/Multi/00611/2013) and FEDER - European
Regional Development Fund through
COMPETE 2020 – Operational Programme Competitiveness and
Internationalization (project: POCI-01-0145-FEDER-006922).}


\abstract{Machine learning techniques have been widely used in attempts to forecast several solar datasets such as the sunspot count, the sunspot area, flare activity, solar wind magnitude and solar storms/coronal mass ejections (CMEs) activity. Most of these approaches employ supervised machine learning algorithms which are, in general, very data hungry. This hampers the attempts to forecast some of these data series, particularly the ones that depend on (relatively) recent space observations such as those obtained by the Solar and Heliospheric Observatory (SOHO) and the Solar Dynamics Observatory (SDO). Here we focus on an attempt to forecast the solar surface longitudinally averaged radial magnetic field distribution (average over absolute values) using a form of spatial-temporal neural networks. Given that the recording of these spatial-temporal datasets only started in 1975 and are therefore quite short, the forecasts are predictably quite modest. However, given that there is a potential physical relationship between sunspots and the magnetic field, we employ another machine learning technique called transfer learning which has recently received considerable attention in the literature. Here, this approach consists in first training the source spatial-temporal neural network on the much longer time/latitude sunspot area dataset, which starts in 1874, then transferring the trained set of layers to a target network, and continue training the latter on the magnetic field dataset. The employment of transfer learning in the field of computer vision is known to obtain a  generalized set of feature filters that can be reused for other datasets and tasks. Here we obtain a similar result, whereby we first train the network on the spatial-temporal sunspot area data, then the first few layers of the neural network are able to identify the two main features of the solar cycle, i.e.\ the amplitude variation and the migration to the equator, and therefore can be used to train on the magnetic field dataset and forecast better than a   prediction
based only on the historical magnetic field data. }


\jnlcitation{\cname{%
\author{Covas E.}} (\cyear{tbc}), 
\ctitle{Transfer Learning in Spatial-Temporal Forecasting of the Solar Magnetic Field}
}

\maketitle

\footnotetext{\textbf{Abbreviations:} MDI, Michelson Doppler Imager; HMI, Helioseismic and Magnetic Imager; SOHO, Solar and Heliospheric Observatory; SDO, Solar Dynamics Observatory; CME, Coronal Mass Ejection; SSIM, Structural SIMilarity Index}

\section{Introduction}
While sunspots have been observed as dark features within the solar disk since ancient times -- some records go back to 800 B.C.\ ancient China \citep{1980AmJPh..48..258S,1989QJRAS..30...59M} -- and have been recorded systematically since the introduction of the telescope in the early 1600s, the actual physical phenomena that presumably originates the sunspots, the solar surface magnetic field, has only been observed consistently in high resolution since the early 1970s, using several ground based and space based observatories. Both datasets now encompass spatial-temporal dimensions, and the sunspot set  is usually depicted in time (Carrington rotation)\footnote{The surface solar rotation varies with latitude and time, meaning any attempt at     tracking features on the Sun's surface over a period of time is obviously a subjective task. Because of this, solar rotation is taken to be 27.2752316 days and each Sun's rotation is given a numeric identifier - the so-called Carrington Rotation Number - starting from number 1 on November 9, 1853.}  versus latitude, the so-called sunspot butterfly\footnote{The ``butterfly wings' pattern is named the butterfly diagram and was first revealed by Edward and Annie Ma\"under in 1904 \citep{1904MNRAS..64..747M}.} diagram \citep{1980SoPh...68..303Y, butterfly}. The equivalent solar surface magnetic field ``butterfly diagram''  is available since 1974, and can be derived from the so-called synoptic magnetograms\footnote{The datasets containing the synoptic magnetograms, which represent the total surface of the Sun in all latitudes and longitudes,  overlap and are sourced from seven solar observatories, each covering different periods of time: Stanford/WSO, NSO/KPVT, NSO/SOLIS, NSO/GONG, SOHO/MDI, UCLA/MWO, and SDO/HMI. For details on calibration and merging of these datasets, see \cite{Riley_2013}.}.

It is generally assumed that the magnetic field originates the sunspot cycle, under the so-called dynamo mechanism (see e.g.\ \cite{9780198512905}, and for reviews see \cite{2003ASPC..286...97O, 2003A&ARv..11..287O, Charbonneau_2010}). This mechanism generates the solar cycle \citep{1844AN.....21..233S}, which, when originally discovered, was thought   to be periodic, with a period close to 11 years. However, it is now thought that the solar cycle is not periodic but has a dynamical behaviour of a low dimensional chaotic nature \citep{2014SSRv..186..525A}.

It is believed that the solar cycle can affect the Earth's magnetic field via what is now called ``space weather'' \citep{Camporeale_2019} and potentially indirectly the Earth's climate  \citep{1991Sci...254..698F, 1995JATP...57..835L, 2006JASTP..68.2053D,9780309265645}. It can also affect human activity, as it is believed that during periods where the solar cycle is at or near its maximum, the increase frequency of solar flares/CMEs can endanger spacecraft's electronics \citep{ 2000AdSpR..26...27W,2011SpWea...9.6001C,2013EGUGA..1510865W}, affect the astronauts' health \citep{2003A&AT...22..861B, 2006GMS...165..367T, 2009SunGe...4...55C,2011AtmEn..45.3806S}, put down electric grids \citep{2005SpWea...3.8C01K}, among other effects. Given these impacts, there is a wealth of research - \cite[for several reviews, see e.g.][and references therein]{1999JGR...10422375H}, and also   \cite{1995SoPh..159..371K, 2003SoPh..218..319U,2012SoPh..281..507P} on forecasting  the solar magnetic activity and/or the sunspot cycle. Predictions are either of a mathematical type (see e.g.\ \cite{1999SoPh..189..217K, 2005ARep...49..495O, 2005SoPh..231..167O}) using only the datasets as a starting point or use e.g.\ dynamo theory with some physical based starting conditions/parameters \citep[see e.g.][and references therein]{2005GeoRL..3221106S, 2005GeoRL..32.1104S, 2006GeoRL..33.5102D}. Other authors use a merge of the two methods -- see e.g.\ \cite{2006GeoRL..3318101H, 2003SoPh..213..203D}. While most solar cycle forecasting focus on the temporal dimension only, there are some examples of attempts to forecast the sunspot butterfly diagram in both latitude and time, i.e.\ spatial-temporal forecast \citep{2011A&A...528A..82J, 2016ApJ...823L..22C, 2014ApJ...792...12M, covas2016, Jiang_2018, covaspeixinhojoao}. There have also been  attempts to simulate the magnetic field \citep{Upton_2014, Upton2018} but these are rarer than sunspot forecasting\footnote{Notice there have been other attempts at forecasting some solar activity features in space and time, e.g.\ flares and CMEs directly \citep{Bobra_2015,Bobra_2016}.}. This is presumably because the magnetic field dataset is quite short, at least when compared with the sunspot spatial-temporal record. 

Neural networks and deep learning have been widely used to forecast the purely temporal aspect of solar cycle and the literature starts in the early 1990s and continues to today \citep[see references in these reviews]{
2012SoPh..281..507P,Pesnell_2016}. Furthermore, neural networks have recently been used to forecast the spatial-temporal dynamics of the sunspot diagram \citep{covaspeixinhojoao}. However, not as much research has been done on forecasting the solar surface magnetic field itself, either by machine learning
approaches or otherwise, particularly in the full  spatial-temporal domain 
\citep[see e.g.][]{
1978GeoRL...5..411S, Mackay2002, Baumann_2004,  2006A&A...459..945S, 2007MNRAS.381.1527J,  
2013PhRvL.111d1106M, Hathaway2016a, Jiang_2018b, 2018MNRAS.479.3791M, Cameron_2018, Bhowmik_2018, Jiang_2018, Jiang_2018_2, 2019arXiv190804474I}. In fact, as far as we are aware, only a few of these articles show a forecast of the magnetic field butterfly diagram, and most research focus on translating the forecast into either a sunspot butterfly diagram, or a pure time series such as the sunspot count or the polar field or the total magnetic field. As far as we can tell, none of these papers show a robust quantification of the goodness of the spatial-temporal forecasts of the magnetic field using a  
numerical quantity that is stable to small changes (e.g.\ translations/rotations) such as the Structural Similarity Index (SSIM) (or the Peak Signal-to-Noise Ratio - PSNR).
 

However, there is a good reason for the scarcity of attempts at forecasting directly the magnetic butterfly diagram: while there is data for the sunspot butterfly diagram consistently since 1874, the magnetic butterfly diagram data starts in 1974, and therefore we have around 13 and an half solar cycles for the former and only four solar cycles for the latter. It is well known that machine learning models used for prediction work better the more data we have. Nonetheless, new techniques have been developed to try to 
overcome this problem. In particular, transfer learning \citep{NIPS1994_959,Yosinski:2014:TFD:2969033.2969197}, a technique whereby one uses a pre-trained neural network on a large dataset (the source or base dataset) to overcome the problem of regression or classification on a smaller but related dataset (the target dataset), has been quite successful. It has been applied extensively to classification problems, e.g.\ object recognition, image classification, and action recognition \citep{Ling_Shao_2015}, sound recognition \citep{2017arXiv170309179C} and other problems \citep[see][for a review]{Pan_2010}. To a lesser extent it has been been applied to forecasting of time series as well \citep{Hu_2016, Qureshi_2017, Ye_2018, Qureshi_2019}. However, when it comes to spatial-temporal data, there are, as far as we are aware, no examples where one applies transfer learning to forecasting an entire spatial-temporal dataset. Notice there are  attempts at using transfer learning in the spatial-temporal domain, but these are slightly different to this article context, involving change detection \citep{Demir_2013,Lyu_2016}, object tracking \citep{Gao_2014}, action detection \citep{Sargano_2017} and image classification \citep{Lv_2018} as opposed to full spatial-temporal forecasting. 
Here we apply the technique of transfer learning to forecasting the longitudinally averaged radial magnetic field strength (i.e. absolute values) by first training a deep neural network on the larger source sunspot area dataset, and then transferring wholly or partially the weights of the trained network to the target network and applying it to the target dataset.

The article is divided as follows. In Section \ref{modelsection} we introduce the model, both the deep neural network and the transfer learning process. In Section \ref{results} we present the results, showing how the transfer learning process improves the forecast of the target dataset, and in Section \ref{conclusions} we draw our conclusions and suggest future research possibilities.







\section{Model}
\label{modelsection}

\subsection{Deep Neural Network}
\label{deepneural network}

Our neural network architecture follows on the approach introduced in \cite{covaspeixinhojoao}, which draws on an technique based on spatial-temporal delays or embeddings \citep{2000PhRvL..84.1890P,Covas,covas2016}. This approach has been empirically demonstrated \citep{Covas_2019} to be optimal or near optimal for forecasting a variety of non-linear spatial-temporal signals. These networks are a generalization of time-delayed neural networks \citep{Waibel:1990:PRU:108235.108263}. The starting point is a spatial-temporal series ${\bf s}$ - a $N$ by $M$ matrix with 
$s^n_m \in \mathbb{R}$, the states of the spatial-temporal series. The embedding vectors ${\bf x}(s^n_m)$ are constructed \citep[see][for details]{covaspeixinhojoao}
 by:
\begin{eqnarray}
\label{embedding}
{\bf x}(s^n_m)&=\{&  s^n_{m-I K }, \ldots,s^n_m, \ldots, s^n_{m+I K},\\
                  && s^{n-L}_{m-I K},\ldots, s^{n-L}_{m},\ldots, s^{n-L}_{m+I K},\nonumber\\
				  && \ldots \nonumber\\ 
                  && s^{n-J L}_{m-I K},\ldots, s^{n-J L}_{m},\ldots, s^{n-J L}_{m+I K} \},
\nonumber
\end{eqnarray}
where $K$ and $L$ represent the spatial and temporal delays or lags, $2I$ is the number of neighbours in space and $J$  
is the number of neighbours in time around the central point $s^n_m$, taken to create the embedding vector ${\bf x}(s^n_m)$. The neural network, a simple deep feed-forward fully connected network, takes as input these embedding vectors of dimension
$(2 I+1)(J+1)$  and as target the value of $s^{n+1}_{m}$ and is then trained by minimizing the local cost function $\mathcal{L}_l$:
\begin{equation}
\mathcal{L}_l = \frac{1}{2 \times \textnormal {batch size}} \sum_{\substack{\textnormal{batch}\\ \textnormal{examples}}}  \norm{s^{n+1}_{m} - f({\bf x}(s^n_m))}^2,
\label{cost}
\end{equation}
where $f({\bf x}(s^n_m))$ is the neural network prediction. The minimization is performed
 using the back-propagation algorithm \citep{ 1986Natur.323..533R,  Lecun98gradient-basedlearning} using
 the Adam optimizer  \citep{2014arXiv1412.6980K} within the well known TensorFlow 
python library \citep{Abadi:2016:TSL:3026877.3026899}.

\subsection{Transfer Learning}
\label{transferlearning}

\begin{figure}[!htb]
\resizebox{0.8\hsize}{!}{\includegraphics{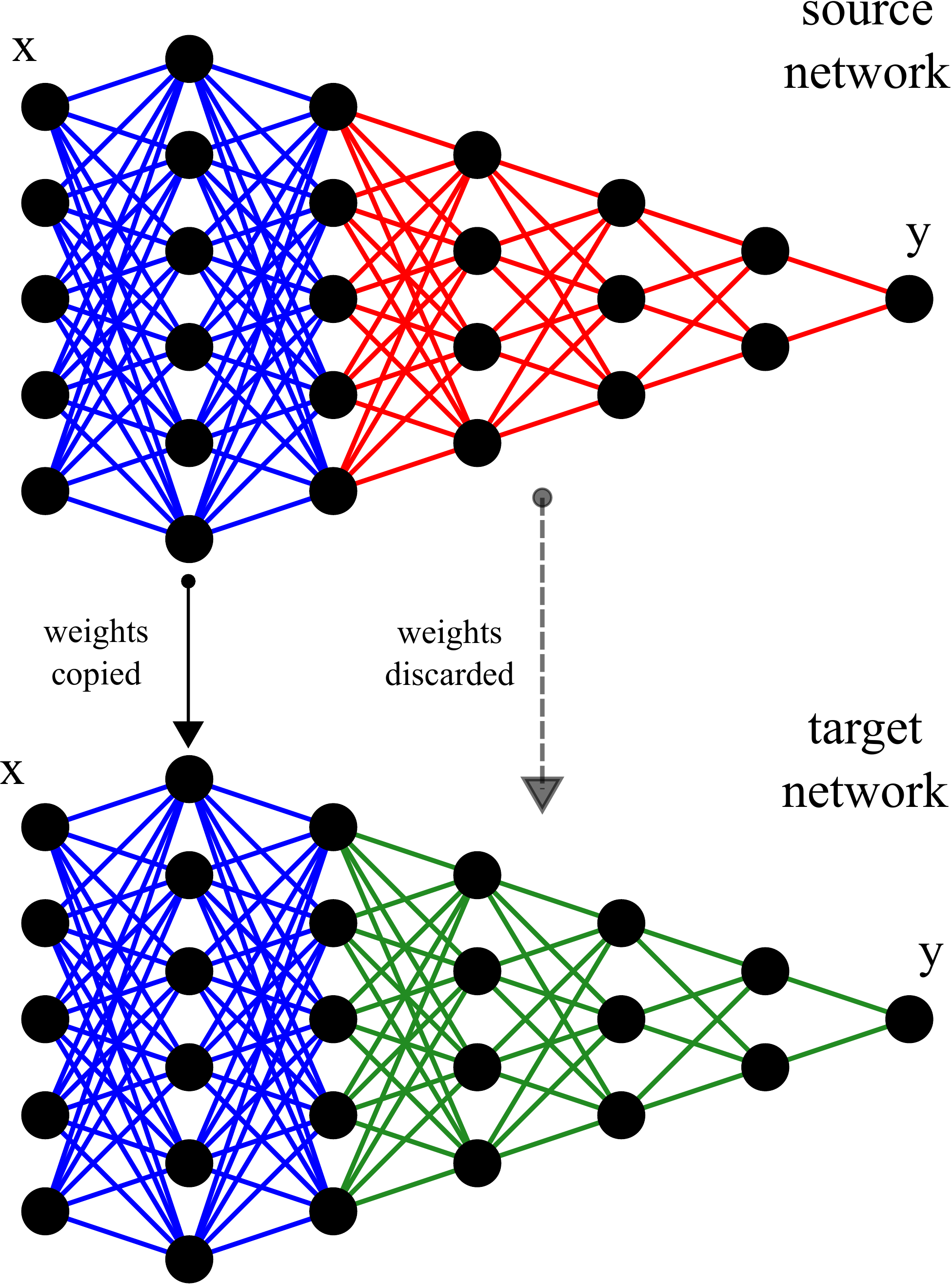}}
\caption{Schematic transfer learning architecture. In this figure $x$ represents the input layer, $y$ the output layer, with the layers in between being the hidden layers. For clarity we used a small amount of nodes in each layer, in reality this article uses a much larger network. 
The source neural network is trained first on the sunspot area data. Then the first few layers 
are copied to the target network (in this figure two layers of weights are copied - in blue), while the other weights (in red) are discarded and new randomly initialized weights are created (in green). The target network is then re-trained on the magnetic field data. There are several options regarding how to treat the copied weights or layers: these can either be frozen, or allowed to be fine-tuned or one can impose a $L^2-SP$ regularization, where the weights $w_i$ are allow to drift but only around the copied source $w_i^0$ values.}
\label{architecture}
\end{figure}

Transfer learning \citep{NIPS1994_959} is a technique whereby one trains a neural network on a large or larger dataset, and then transfer part or the whole set of layers/trained weights to another neural network which is then re-trained on a small or smaller dataset. The approach is used for similar or related datasets and the intuition is that, once trained, the higher layers have acquired the ability to detect generic features and this will be useful on the subsequent task. In Figure \ref{architecture} we show a schematic of this technique, where the higher or top layers (the generic layers) are transferred and where the lower or bottom layers (the specific layers) are re-initialized with random weights. One can freeze, i.e.\ stop the back-propagation algorithm  at the last of those transferred layers or reuse those weights and fine-tune them to the new task  \citep{Yosinski:2014:TFD:2969033.2969197}. A third and intermediate approach has been developed \citep{L2SP} where one introduces an explicit bias in the form of a $L^2$ type of regularization towards the copied weights or starting points (hence the coined term $L^2-SP$). Formally, the $L^2$ regularization consists of  a term $\omega(w)$ added to the cost function \eqref{cost} (which is minimized by the back-propagation algorithm) of the form
\begin{equation}
\omega(w)= \frac{\beta_1}{2} \sum_{i \in \textnormal{layers}} \norm{w_i}^2,
\label{L2_equation}
\end{equation}
where $w \in \mathbb{R}^n$ represent all the weighs on the target network
and the intensity of the penalty is controlled by $\beta_1$, the regularization parameter. The $L^2-SP$ regularization term
\citep[see][for details]{L2SP} is given by 
 \begin{equation}
\omega(w)= \frac{\beta_1}{2} \sum_{i \in \, \substack{\textnormal{copied}\\ \textnormal{layers}}} \norm{w_i-w^0_i}^2 + 
\frac{\beta_2}{2} \sum_{i \in \, \substack{\textnormal{other}\\ \textnormal{layers}}} \norm{w_i}^2,
\label{L2SP_equation}
\end{equation}
where $w^0$ represents the vector of all transferred weights of the trained source neural network (the {\it blue} connecting weights in Figure \ref{architecture}), while  $\beta_1$ represents the penalty on the transferred generic weights and $\beta_2$  the penalty on the new specific weights. In this article we will set, for simplicity, $\beta_1=\beta_2$ and refer to $\beta_1$ and  $\beta_2$ as  $\beta$. 

\section{Results}
\label{results}

\subsection{Dataset and parameters}


We take the publicly available sunspot area data
\footnote{We use the dataset publicly available in \url{http://solarcyclescience.com/AR_Database/bflydata.txt } \citep{butterfly}.}. 
The dataset we use contains the area of the Sun covered by sunspots from 
Carrington Rotation 275 (around April 1874) to Carrington Rotation 2195 (approximately September 2017), and is given in units of millionths 
of a hemisphere. The data is spaced in 50 latitude bins distributed uniformly in $\sin(\theta)$, where $\theta$ is the latitude. For training the 
network, we use the first 1802 Carrington Rotations, from the number 275 to the number 2076, i.e.\ we use cycle 11 to 23. For this dataset  the optimal values for spatial/temporal lags and the number of spatial/ 
temporal lags to be used were, respectively: $I^*=2$, $J^*=6$, $K^*=9$ and $L^*=70$, as calculated in \cite{covas2016} and shown to be optimal under 
a variety of stress scenarios in \cite{Covas_2019}.
 
The longitudinally averaged magnetic field strength data, calculated by averaging the absolute value of the field  on each synoptic map 
over all longitudes, is publicly available and can be derived from the   synoptic magnetogram maps datasets, as long as one 
uses the correct conversion functions between the different datasets for calibration \citep{Riley_2013}. Our data was kindly provided by David H.\ 
Hathaway from NASA (private communication). For the avoidance of doubt, what we used was the average across all longitudes of the absolute value of the radial photospheric magnetic field, not the absolute value of the average across longitudes of the field. The data encompasses Carrington Rotation 1623 to 2195 and already contains all the necessary calibration and merging of 
the several instruments' datasets into one single consistent file, together with the longitudinal averaging after setting all field values to its absolute value. This magnetic field set has 180 latitudinal slices (while the 
sunspot area data has 50 slices). There were also some time slices that were not available, i.e.\ they were zero for all 
latitudinal bins, this is due to technical problems with the instruments within those time periods. For these missing slices we simply extended the data 
constantly from the previously available one and avoided trying to guess the intermediate values by interpolation/extrapolation.
 
 For consistency with previous work, we kept all parameters for both source and target network as in \cite{covaspeixinhojoao}, except the following absolutely necessary modifications. First we use a deeper neural network that in \cite{covaspeixinhojoao}. We settled on 5 hidden layers (a network with a total of 7 layers if we count the input and the output layers) by analysing the performance of the sunspot forecast, via the SSIM value\footnote{The SSIM index is widely used in computer vision  \citep{Wang04imagequality}. It is a numerical quantity with values 
 $\textnormal{SSIM} \in [0,1]$ and a value of $\textnormal{SSIM}=1$ occurs when one calculates it between two  identical images or datasets. In our case study here, a $\textnormal{SSIM}=1$  corresponds to a perfect spatial-temporal forecast.}, as a function of the number of hidden layers.  We used  the following number of nodes in each consecutive hidden layer: $n_{h_1}=70$, 
$n_{h_2}=60$, $n_{h_3}=50$, $n_{h_4}=40$ and $n_{h_5}=30$.  As in \cite{covaspeixinhojoao}, we use the Adam optimizer algorithm 
 for faster convergence, with a early stop which is defined by a sufficient slowdown of the decay of the global cost function on the entire training set. This gave us \SI{267000} epochs or iterations of the Adam optimization for the sunspot data, and \SI{192000} epochs for the magnetic field data (we calculate the global cost every \SI{1000} iterations as it is quite costly to calculate that value in the entire training set).
 Second, as the magnetic field data is much shorter in time, the value of $J=J^*=6$ as calculated in \cite{covas2016} cannot be used. This is because we would have to  construct an embedding vector
 $ {\bf x}(s^n_m)$ with $J+1=7$ back in time neighbours (see Equation \eqref{embedding}). However, given the optimal time delay lag, $L^*=70$,  we would have to use data back in $(J+1)L=490$ Carrington Rotations,
 which is almost the amount of slices we have on the magnetic field data. We believe, without loss of generality, as we are reducing the amount of information for training, that we can use a lower and sub-optimal value of $J$. We pick $J=4$ as the highest we can go without compromising the amount of necessary embedding vectors to achieve convergence of the global cost function within the training set. Third, we have to scale the spatial lag $K$ since we have 180 latitudinal slices in the magnetic field data and 50 on the sunspot area data. So the optimal spatial lag for the magnetic field embedding vector construction $K_\textnormal{mag}$ is given by:
  \begin{equation}
 K_\textnormal{mag}=
 \floor*{180/50\times K^*} = \floor*{180/50\times 9} = 32,
 \end{equation}
 where $K^*=9$ is the optimal spatial lag for the sunspot data  as demonstrated in \cite{covas2016}.
 The rest of the hyper-parameters of the neural network were taken as in \cite{covaspeixinhojoao}: a mini-batch size of one (so we use stochastic gradient descent learning), a logarithmic normalization of the inputs scaled
 with $x \to \alpha_{nor}+\nicefrac{\ln(1+x)}{\beta_{nor}}$, where $x$ represents the initial embedding vectors,
and $\alpha_{nor}$ and $\beta_{nor}$ are the arbitrary shift and scaling constants, we took $\alpha_{nor} = 10$ and $\beta_{nor} = 0$; we used weight initialization with random values taken from a constant distribution between $[0,1]$ and shifted by $\alpha_{rng}$ and scaled
by $\beta_{rng}$, we took $\alpha_{rng} = 10^{-2}$ and $\beta_{rng} = -0.5$. We, as in \cite{covaspeixinhojoao} used the logistic sigmoid function as the activation on all layers.
There was one more final modification, this time only applied only on the target network. Since the maximum of the original magnetic field data was different from the maximum of the sunspot data, we therefore used a different normalization scaling $\beta_{nor}\approx 5.61843$ for the magnetic field data, so that the maximum/minimum of the normalized data on both the source and target network were the same.

\subsection{Results}

\begin{figure}[!htb]
\resizebox{0.9\hsize}{!}{\includegraphics{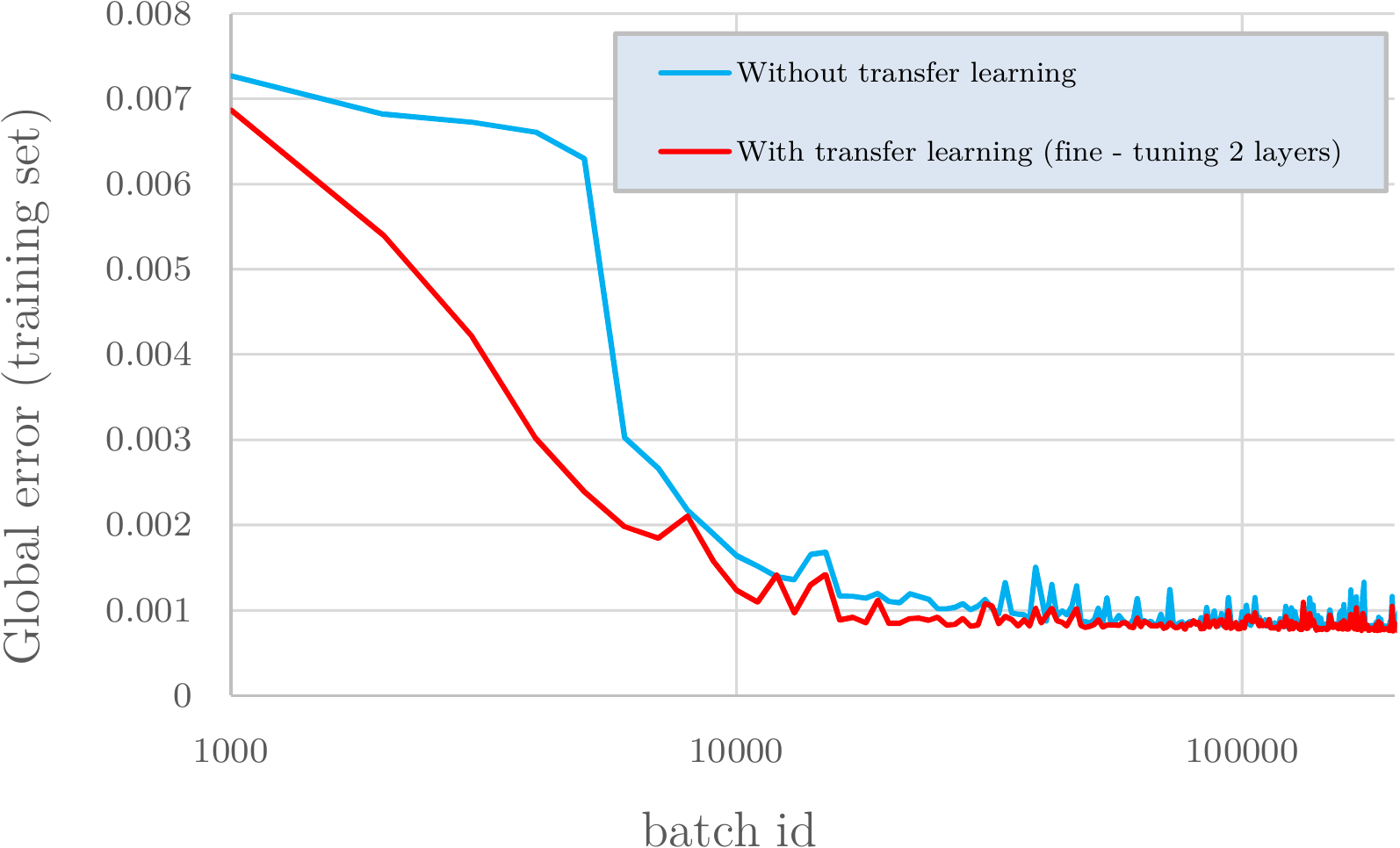}}
\caption{Global training cost $\mathcal{L}_g$ (see Equation \eqref{global_cost}) for the magnetic field dataset.
This example shows the global cost on a run without transfer learning (random weight initialization) 
and a run where the weights were transferred up to the second hidden layer, i.e.\ two sets of weights were transferred. No layer freezing was applied, so this is an example of fine-tuning. The result shows how  transfer learning can potentially help, first by having already a lower cost at the starting phase, second by having a sharper slope towards convergence, and third by having a lower asymptote value.}
\label{Errorfunction}
\end{figure}

Figure \ref{Errorfunction} depicts our first result, showing the global error as measured by the 
global cost function $\mathcal{L}_g$ calculated using the whole training examples (magnetic field dataset)
\begin{equation}
\mathcal{L}_g=\frac{1}{2}  \sum_{\substack{\textnormal{training}\\ \textnormal{examples}}} \norm{y^{\textnormal{\small{pred}}}-y}^2,
\label{global_cost}
\end{equation}
where $y^{\textnormal{\small{pred}}}$ is the predicted value of the neural network, while $y$ is the target value. The result shows that the 
error decays/converges faster when one applies transfer learning.

\begin{figure*}[!htb]
\resizebox{\hsize}{!}{
\includegraphics{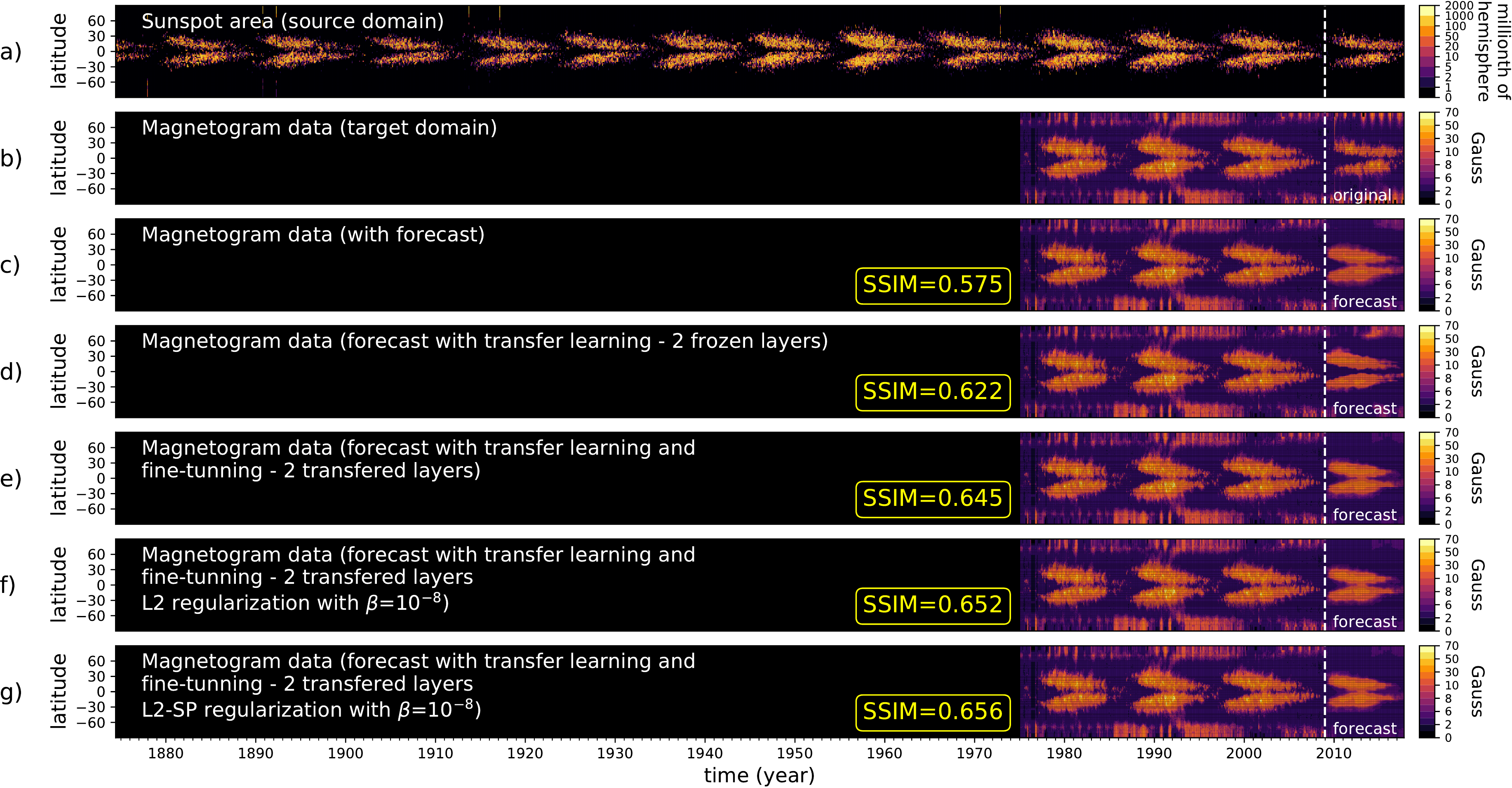}}
\caption{Results of transfer learning against the pure forecast without transfer learning for the the  magnetic field dataset. 
In panel (a) we show the source domain, with the sunspot data from 1874 to 2017. In panel (b) we show the target domain, with the
longitudinally averaged magnetic field strength from 1974 to 2017 (courtesy D.\ Hathaway).
In panel (c) we show the pure forecast, using the target neural network without any transfer learning, based on the training set encompassing the period
December 1974 (Carrington Rotation 1623) to October 2008 (Carrington Rotation 2076) and the test or forecast set being the data
within the period November 2008 (Carrington Rotation 2077) to September 2017 (Carrington Rotation 2195). 
In panel (d) we train the source network using the sunspot dataset (from Carrington Rotation 295 to 2076). The weights vectors, the first two layers, are then transferred to the target network, and the first two hidden layers are then frozen. The result is an improvement, both visually and in the SSIM value, which increases considerably. In panel (e) we show the results for fine-tuning, and as suggested in the literature, there is a further improvement, and although not easy to see visually, it is measured by the SSIM value, which increases again. Following the results of \cite{L2SP}, we test the forecast using fine-tuning but with a $L^2$ regularization with $\beta=10^{-8}$ - panel (f) and
a $L^2-SP$ regularization, again with $\beta=10^{-8}$ - panel (g), and obtain even better results, measure by an improvement, albeit small, in the SSIM value. Note the colour scale is deliberately non-linear to enhance the contrast of the images.
}
\label{results_transfer_learning}
\end{figure*} 

In Figure \ref{results_transfer_learning} we show the main results of this article. It depicts the forecasting without transfer learning - shown in panel (c), against several approaches to transfer learning, namely: transfer $n$ (in this Figure, $n=2$) hidden layers set of weights with a subsequent freezing of those layers - shown in panel (d); transfer two hidden layers set of weights with a subsequent fine-tuning - panel (e); transfer two hidden layers set of weights with a subsequent fine-tuning and a $L^2$ regularization applied to all layers - panel (f); and finally transfer two hidden layers set of weights with a subsequent fine-tuning and a $L^2-SP$ regularization applied to all layers - panel (g). The results seem to show, both visually and in terms of the SSIM index, that we have an improvement in the predictions as we apply transfer learning and that even further improvements are possible with more sophisticated forms of transfer learning. Notice that in all the runs for predictions of the magnetic field - (c) to (g) - we used the same number of iterations of the optimizer: \SI{192000} epochs. This ensures that all forecasting approaches are in the same footing in terms of the neural network architecture, convergence and hyper-parameters, and that the only difference is really the transfer learning methodology, which is what we want to test in this article.


Some differences can also be identified between the observed and the predicted
magnetic field butterfly diagrams, e.g.\ the observed real data has a more grainy structure, and the predicted equatorial asymmetry is not as expected. Nonetheless, the results show that, first, one can reproduce the qualitative aspects of the solar cycle (the amplitude variations and the migration towards the equator) and second, that transferring the weights seems to improve the  magnetic field forecast. In fact it is interesting to remark that the cycle 24 magnetic field latitudinal average amplitude is around half of the amplitudes for cycles 21-23, so the neural network is able to extrapolate reasonably well outside its training set, which consists of three stronger solar cycles. We also note that transferring without freezing the copied layers (fine-tuning) is more effective that freezing them. Furthermore using a $L^2$ regularization on all layers or even better a $L^2-SP$ regularization improves the forecast. This improvement, if not entirely visually obvious  in Figure \ref{results_transfer_learning}, at least can be seen clearly in the numeric SSIM index results. This is consistent with the results of \cite{Yosinski:2014:TFD:2969033.2969197} and of
\cite{L2SP}.

\begin{figure}[!htb]
\resizebox{0.9\hsize}{!}{
\includegraphics{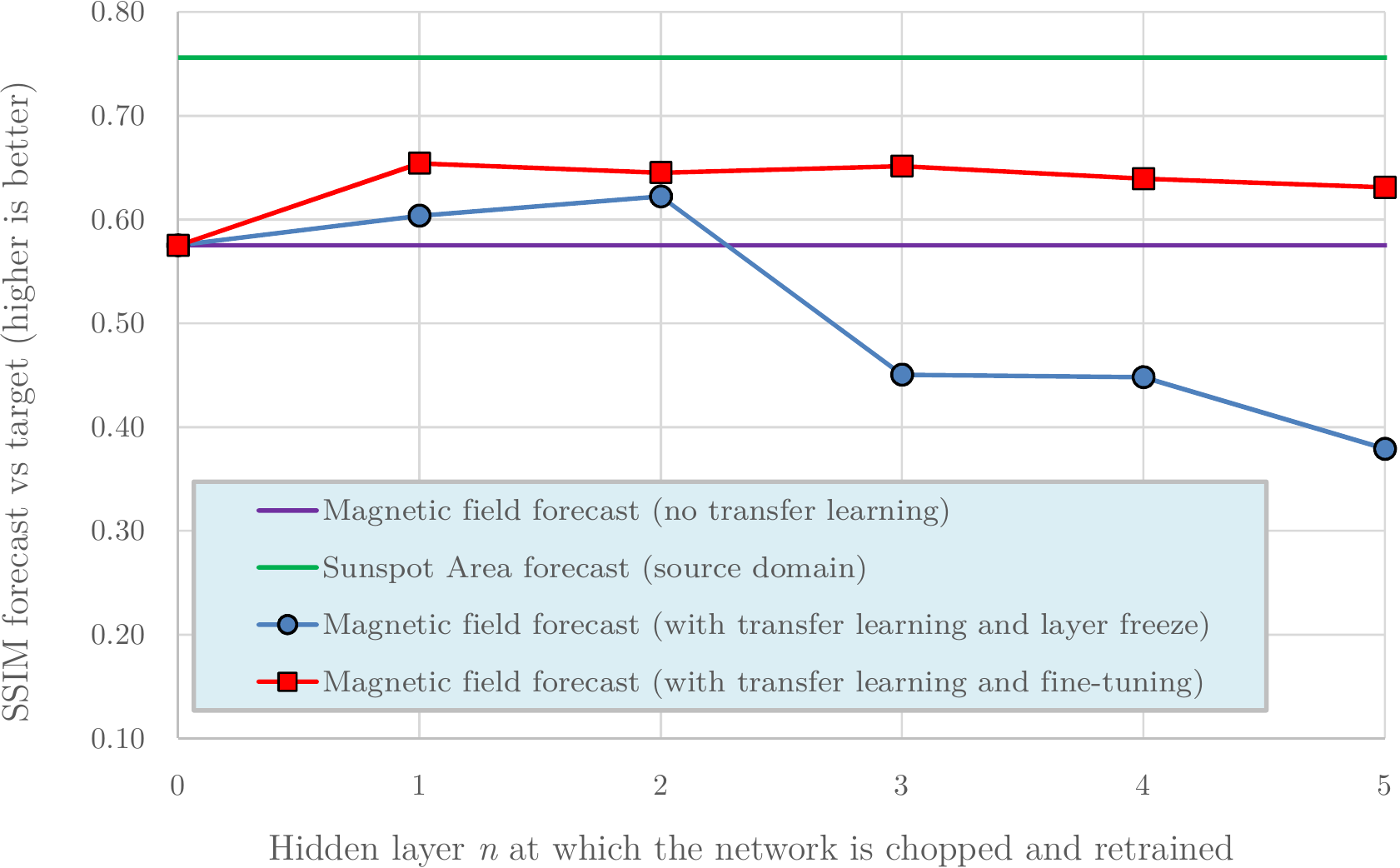}}
\caption{The SSIM index of the forecast against the target data. All parameters are the same across the figure except the hidden layer index $n$, which represents the layer at which the source network is chopped and copied to the target network, after which the target network is then retrained. It also shows, for comparison, the source SSIM index for the sunspot area data forecast (in green) and the SSIM index for the pure forecast without transfer learning for the magnetic field (in purple). The results are very similar as the one obtained in \cite{Yosinski:2014:TFD:2969033.2969197}. We see (blue filled circles) that the performance initially increases from the base comparison, but then drops quite a lot after layer $n\ge 2$ due to representation specificity when we freeze the layers, while for transfer plus fine-tuning (red filled squares), performance increases against the base comparison across all values of $n$ and so it seems to improve generalization.
}
\label{SSIMversusNumberFrozenWeightLayers}
\end{figure}

Following on the results of \cite{Yosinski:2014:TFD:2969033.2969197}, where the authors address the question on how transferable are features in deep neural networks,  we also analyse the effectiveness of the transfer learning, as measured by the SSIM index, as a function of the number of the layers transferred from the source network to the target network. These results are depicted in Figure \ref{SSIMversusNumberFrozenWeightLayers}, which compares directly to Figure 2 in \cite{Yosinski:2014:TFD:2969033.2969197}. It shows that, in general, transfer learning can help to improve the prediction. The fine-tuning approach, whereby the weights are transferred up to a certain layer and then the back-propagation algorithm is allowed to update all weights, is better than the transfer learning literature's original approach whereby the weights are only updated up to a certain layer: the frozen layers approach.

\begin{figure}[!htb]
\resizebox{0.9\hsize}{!}{
\includegraphics{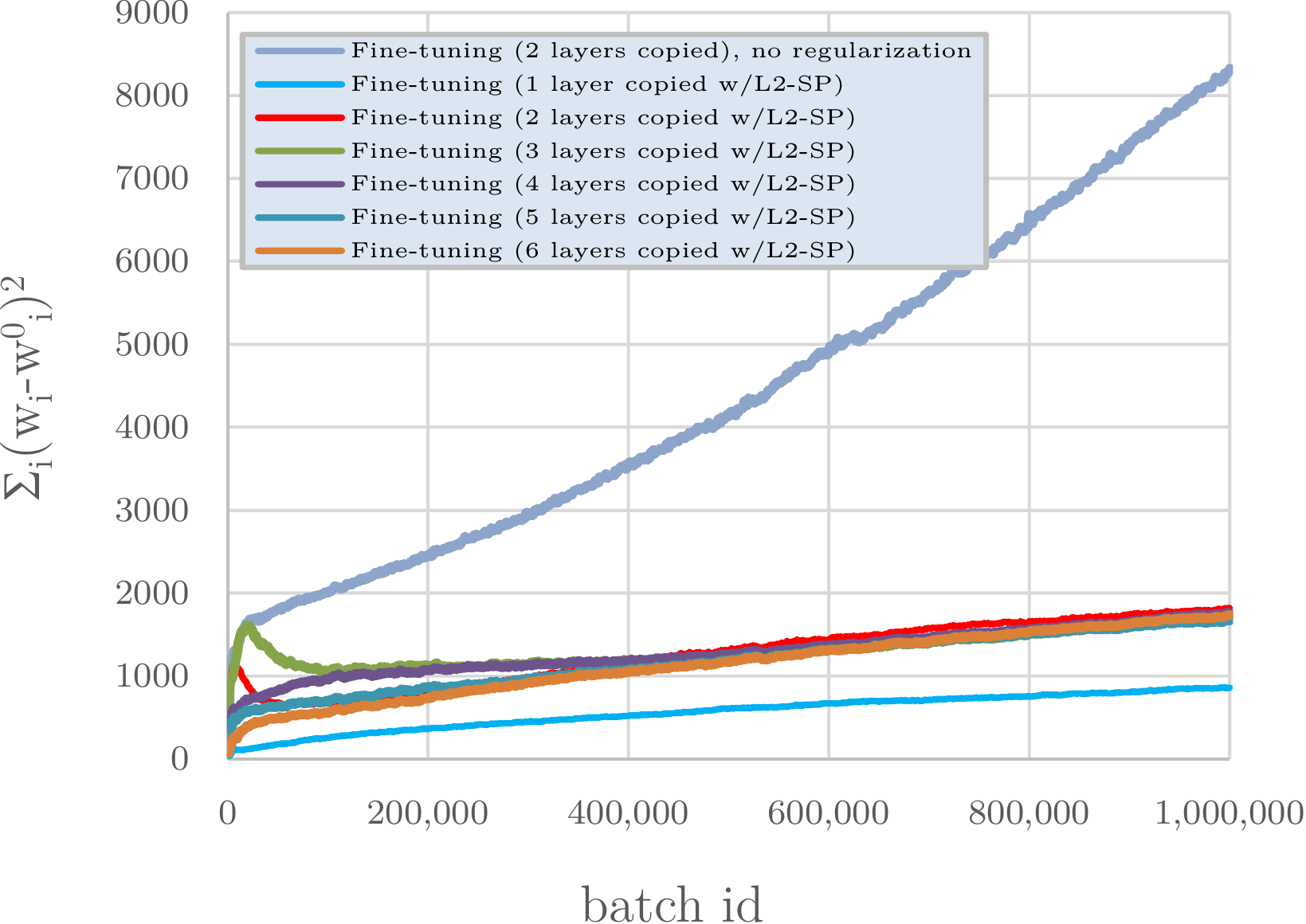}}
\caption{The sum of the square of differences $\Delta w $ as in Equation \eqref{sumDeviationWeightsFormula} for transfer learning with 
fine-tuning. One can see that the $\Delta w $ diverges pretty quickly for the case without regularization (we show the case for two copied layers), while $\Delta w $ 
for $L^2-SP$ regularization  (with $\beta=10^{-8}$) is more stable, independently of the number of  copied layers.
}
\label{sumDeviationWeights}
\end{figure}

Finally, following on the results of \cite{L2SP}, we analyse how a proposal for an intermediate approach between the extreme of freezing the copied layers and the other extreme of allowing the optimizer to fine-tuning freely, the so-called $L^2-SP$ (and the $L^2$) regularization can help to improve the accuracy of the prediction. The $L^2-SP$ regularization approach has been introduced to overcome a problem with transfer learning, i.e.\ if we freeze the copied layers then while the general part of the network, responsible for feature detection, is kept intact, this is too rigid and performance can drop due to different representation specificity between the two domains, as seen in Figure \ref{SSIMversusNumberFrozenWeightLayers} (blue filled circles). If we do not freeze the copied layers but allow fine-tuning then another problem can show up, that the weights or feature detectors in the copied layers as modified by the optimizer start to diverge too much away from the source representation, and we do not retain much of the knowledge or the features learned on training within the source domain. In other words, on the former approach we keep too much ``prejudice'' and in the latter approach we ``forget'' the general feature representation we learned. In \cite{L2SP}, the authors investigated several regularization approaches that explicitly restrain the weights to stay close to the initial solution from the source domain. They showed that using an explicit inductive bias via a modified $L^2$ type of regularization towards the initial solution or weights can help, and that within many possible functional forms of regularization, that the $L^2-SP$ as in Equation \eqref{L2SP_equation} was the best across the spectrum and the simplest one to use. 
Given their results, we first investigated if indeed the fine-tuned weights $w_i$ were diverging too fast from the copied weights $w_i^0$ by plotting the sum of the square of differences $\Delta w$ as:
\begin{equation}
\Delta w  \triangleq \sum_{i \in \, \substack{\textnormal{copied}\\ \textnormal{layers}}} \norm{w_i-w^0_i}^2.
\label{sumDeviationWeightsFormula}
\end{equation}  
The result, for  transfer learning with 
fine-tuning and two copied layers is shown in Figure~\ref{sumDeviationWeights} against fine-tuning with $L^2-SP$ regularization applied to the cost function (with $\beta=10^{-8}$). It clearly shows that $\Delta w $ diverges pretty quickly as the learning process evolves (i.e.\ we forget the general feature representation). However, if one  applies a regularization to keep the weights within the neighbourhood of the initial copied weights $w_i^0$, then $\Delta w $ seems to stabilize asymptotically, as we desire. 

\begin{figure}[!htb]
\resizebox{0.9\hsize}{!}{
\includegraphics{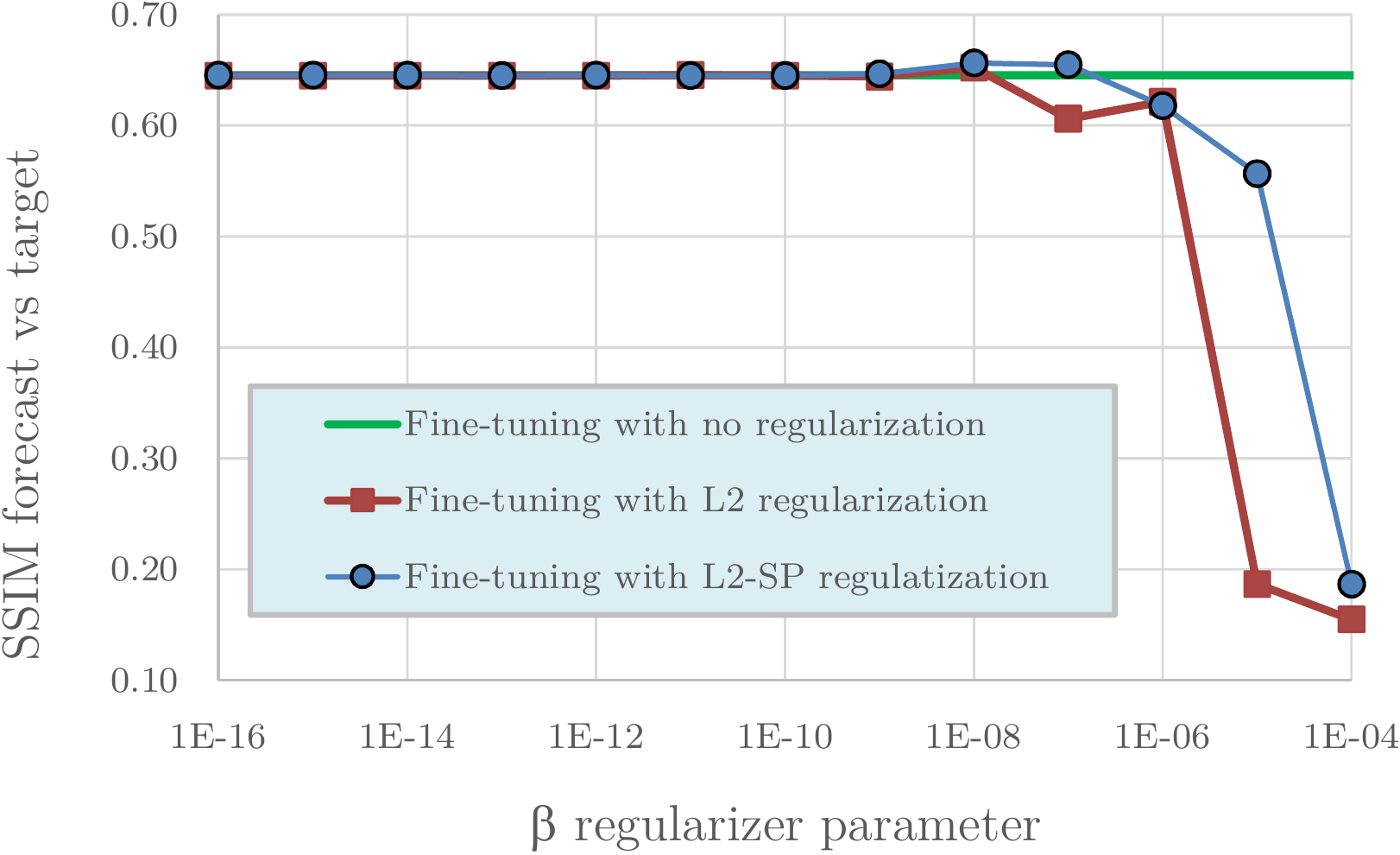}}
\caption{The SSIM index of the forecast versus target data for no regularization (green line), $L^2$ regularization (red filled squares) and $L^2-SP$ regularization (blue filled circles) as a function of the regularization penalty parameter $\beta$ for fine tuning with two copied layers.
}
\label{SSIMVersusBetaRegularizer}
\end{figure}

Furthermore, as already shown in Figure \ref{results_transfer_learning} on panel (g), there is an improvement if we implement this penalty function towards the 
initial copied weights $w_i^0$. We note that there is also an improvement if we implement just a plain $L^2$ regularization - panel (f), just not the maximum improvement we manage to obtain in this study. To clearly show this, we plot in Figure \ref{SSIMVersusBetaRegularizer} the SSIM of the forecast against the target versus the strength of the penalty, as measured by the regularization parameter $\beta$. It shows, as expected, that the inclusion of the constraint has no effect for very low values of $\beta$, and correspondingly, that for high values it destroys any chance of a good forecast, as it does not allow any learning to occur during the training phase. A sweet spot interval exists around $\beta \in 
\left[ 10^{-8} - 10^{-6}\right]$ and within it the $L^2-SP$ regularization outperforms the basic $L^2$ regularization, again as expected.

\section{Conclusions}
\label{conclusions}

We use the technique of transfer learning to enhance the performance of the spatial-temporal forecast of the  longitudinally averaged solar surface magnetic field. As the length of the data is quite short (around 4 solar cycles - just under 43 years of data), it is quite difficult to forecast with a reasonable precision, and we can only obtain a modest value of the SSIM accuracy index for a pure forecast with a neural network. However, if we use the sunspot area dataset, available for longer (just over 13 solar cycles -  over 143 years) to first train, and then transfer the whole or part of the weights to the target network, we can enhance the forecast, and obtain higher values of the SSIM index.
As far as we are aware, this is the first time that the forecast of the spatial-temporal solar surface magnetic field has been attempted using neural networks and also using transfer learning. 
The results show that the approach can reproduce qualitatively some of the main features of the spatial-temporal dynamics of the surface magnetic field evolution diagram, 
such as the overall cycle amplitude modulation, and the cycle migration to the equator.
We have explored several approaches for transfer learning  and found that while the basic weight transfer approach improved the accuracy of the forecast with regards to no transfer learning based predictions, that the accuracy could be further enhanced by using a novel approach called $L^2-SP$ regularization, whereby a  penalty is used to force the transferred weights not to deviate too far from the original source domain values.

We believe that this research points in the right direction in terms of using longer solar cycle related datasets in order to improve the accuracy and precision of the forecasts for shorter datasets. While some of the data goes back 400+ years (the sunspot count data) and is thought of being the
longest continuously recorded measurement made in science \citep{2013Natur.495..300O}, other data such as the solar surface magnetic field intensity in both latitude and longitude, the 530.3 nm green coronal time series and the 10.7 cm radio flux data \citep{2015SoPh..290.3095B}, proxies such as the geomagnetic {\it aa} indices \citep{1972JGR....77.6870M,1993GeoRL..20.2703N} and others have only started being recorded with the advent of more advanced ground based and space based observatories and therefore the technique of transfer learning could help to improve those predictions attempts.
These forecasts are not only relevant from the pure scientific point of view, but also because predicting solar events is now of critical importance in the emerging field of  space weather.





\backmatter

\section*{Acknowledgments}

We would like to thank Reza Tavakol from Queen Mary University of London for useful conversations on forecasting. We specially like to thank
David Hathaway from NASA's Ames Research Centre for providing the solar surface longitudinally averaged total magnetic flux distribution (absolute values) and the sunspot area data on which all the results of this article are based upon.
We also thank In\^{e}s Nolasco from Queen Mary University of London for insightful discussions on the subject transfer learning.
CITEUC is funded by National Funds through \fundingAgency{FCT - Foundation for Science
and Technology} (project: \fundingNumber{UID/Multi/00611/2013}) and \fundingAgency{FEDER - European
Regional Development Fund} through
COMPETE 2020 – Operational Programme Competitiveness and
Internationalization (project: \fundingNumber{POCI-01-0145-FEDER-006922}).
The Wilcox Solar Observatory is supported by NASA. The SOLIS, GONG and
KPVT programs are managed by the NSO, which is operated by AURA, Inc. under a
cooperative agreement with the NSF. SOHO/MDI is a project of international cooperation
between ESA and NASA. The MWO data is from the synoptic program at
the 150-Foot Solar Tower of the Mt. Wilson Observatory. The Mt. Wilson 150-Foot Solar Tower is operated
by UCLA, with funding from NASA, ONR and NSF, under agreement with the Mt. Wilson Institute. HMI
data are courtesy of the JSOC Science Data Processing team at Stanford
University.

\subsection*{Author contributions}

All authors listed, have made substantial, direct and intellectual contribution to the work, and approved it for publication.

\subsection*{Financial disclosure}

None reported.

\subsection*{Conflict of interest}

The authors declare that the research was conducted in the absence of any commercial or financial relationships that could be construed as a potential conflict of interest.


\bibliography{eurico}

\end{document}